\documentclass[12pt]{iopart} %\newcommand{\gguide}{{\it Preparing graphics for IOP journals}}
\usepackage{hyperref}
\usepackage{indentfirst}
\usepackage[latin1]{inputenc}
\usepackage{cite}
\usepackage{iopams}

\begin{document}

\title{Gravastar model in  Randall-Sundrum braneworld}
\author{Jos\'e D V Arba\~nil$^1$, Pedro H R S Moraes$^{2,3}$ and Manuel Malheiro$^2$} 
\address{$^1$Departamento de Ciencias, Universidad Privada del Norte, Avenida Guardia Peruana 890 Chorrillos, Lima,  Peru\\
$^2$Departamento de F\'isica, Instituto Tecnol\'ogico de Aeron\'autica, Centro T\'ecnico Aeroespacial, 12228-900 S\~ao Jos\'e dos Campos, S\~ao Paulo, Brazil\\
$^3$UNINA - Universit\`a degli Studi di Napoli Federico II - Dipartamento di Fisica - Napoli I-80126, Italy}
\eads{jose.arbanil@upn.pe,moraes.phrs@gmail.com,malheiro@ita.br}

\begin{abstract}
In this work we derive a gravastar model in Randall-Sundrum II braneworld scenario. Gravastars (or gravitationally vacuum stars) were proposed by Mazur and Mottola as systems of gravitational collapse alternative to black holes. The external region of the gravastar is described by a Schwarzschild space-time, while its internal region is filled by dark energy. In between there is a thin shell surface with ultrarelativistic matter (sometimes referred to as stiff matter). We obtain  solutions for some physical quantities of the braneworld gravastars. Those are compared to original Mazur-Mottola results as well as with some gravastar solutions in alternative gravity theories. The consequences of the braneworld setup in gravastar physics is deeply discussed.   
\end{abstract}
\pacs{04.40.Dg, 04.20.-q, 04.20.Cv}\maketitle

\section{Introduction}\label{sec:int}

The Randall-Sundrum II braneworld model (RSII, for short) describes the Universe as a single brane embedded in a five-dimensional (5D)  nonfactorazible background geometry \cite{randall/1999}. It reproduces properly the Newtonian and General Relativity theories of gravity in the referred regimes \cite{maartens/2004,garriga/2000,figueras/2011,kim/2004}. 

RSII has been applied to different areas, yielding remarkable results. Several aspects of RSII cosmology were investigated, for example, in References \cite{ramirez/2004,holanda/2013,hebecker/2001,meehan/2014}. Some constrains to the braneworld quantities have been put from different approaches \cite{tsujikawa/2004,liddle/2003,yagi/2011,mm/2014}.

The astrophysics of stellar objects was also deeply analysed in RSII. C. Germani and R. Maartens have firstly shown that in RSII the vacuum exterior of a spherical star is not in general a Schwarzschild space-time, but presents radiative-type stresses generated by 5D graviton effects \cite{germani/2001}. In \cite{visser/2003}, the 4D Gauss and Codazzi equations were solved for an arbitrary static spherically symmetric star. It was shown how the 4D boundary data should be propagated into the 5D bulk in order to get the full space-time geometry. Further properties of compact stars in RSII were studied in \cite{la/2017}. It was found a new branch of stellar configurations that can violate the general relativistic causal limit and that may have an arbitrarily large mass. Moreover, the properties of quark and hadronic stars were analysed in \cite{la/2015}.

RSII has also been widely applied in the astrophysics of black holes (BHs), also yielding remarkable results. For instance, the properties of gravitational lensing by BHs in RSII were explored in \cite{bin-nun/2010}.  In Reference \cite{wang/2016}, the authors studied the process of gravitational collapse driven by a massless scalar field which is confined to the brane. Further BH analysis in RSII may be checked in References \cite{abdolrahimi/2013,abdolrahimi/2013b,tanahashi/2008}. 

In the year of 2016, the first detection of gravitational waves was reported \cite{abbott/2016} by the Advanced LIGO (Laser Interferometry Gravitational Wave Observatory) Team. It was claimed that the detected gravitational wave sign was generated at a redshift $z\sim0.09$ by a BH binary system. Later in the same year, it was argued that the signal-to-noise and quality of the referred data were such that there was some room to alternatively interpret such an event, as a gravastar (gravitationally vacuum stars) binary system \cite{chirenti/2016}.

Gravastars were proposed in \cite{mazur/2004} by Mazur and Mottola as systems of gravitational collapse which are alternative to BHs. The external region of a gravastar is described by a Schwarzschild space-time, such that $p=\rho=0$, with $p$ and $\rho$ being the pressure and matter-energy density, respectively, whereas its surface is a thin shell of ultrarelativistic matter, with $p=\rho$. Its internal region is filled by dark energy, with $p=-\rho$. 

In \cite{chirenti/2007}, it was shown that it is possible to discern a gravastar from a BH of the same mass due to their different quasi-normal modes. The problem of ergoregion instability for the viability of gravastars was also investigated in \cite{chirenti/2008,cardoso/2008}. In \cite{harko/2009}, the possibility of distinguishing BHs and gravastars using the properties of their accretion disks was considered. Observational constraints were put in gravastars from well-known BH candidates \cite{broderick/2007}. Further observational distinguishment between BHs and gravastars were discussed in \cite{sakai/2014}, in which it was argued that high-resolution very-long-baseline-interferometry observations can contribute on this regard in near future.

Moreover, it should be remarked that by means of the usual Tolman-Oppenheimer-Volkoff equation, it was shown that gravastars material content cannot be described by perfect fluids \cite{cattoen/2005}. Instead, they should have anisotropic pressures.   

On this regard, it is well-known that the 5D set up of RSII and other braneworld models induce anisotropy in brane objects, such that it might be interesting and valuable to investigate gravastars in the braneworld. In the present paper we will obtain and investigate gravastars solutions in RSII.

\section{Braneworld field equations}\label{sec:bw}

\subsection{Basic equations}

The field equations of RSII in terms of an effective energy-momentum tensor read \cite{germani/2001,shiromizu/2000,maartens/2000}
\begin{equation}\label{bw1}
G_{\mu\nu}=k^{2}T_{\mu\nu}^{\rm eff},
\end{equation}
with $k^{2}=8\pi G$, $G$ is the newtonian gravitational constant, the speed of light $c=1$ and such that the effective total energy density and pressure, anisotropic stress and energy flux read, respectively:
\begin{eqnarray}
&&\rho^{\rm eff}=\rho+\frac{1}{2\lambda}\left(\rho^{2}+\frac{12}{k^{4}}\mathcal{U}\right),\label{bw2} \\
&&p^{\rm eff}=p+\frac{1}{2\lambda}\left[\rho(\rho+2p)+\frac{4}{k^{4}}\mathcal{U}\right],\label{bw3} \\
&&\pi_{\mu\nu}^{\rm eff}=\frac{6}{k^{4}\lambda}\mathcal{P}_{\mu\nu},\label{bw4} \\
&&q_\mu^{\rm eff}=\frac{6}{k^{4}\lambda}\mathcal{Q}_\mu,\label{bw5}
\end{eqnarray}
with $\lambda$ being the brane tension and the bulk cosmological constant was taken such that the brane cosmological constant is null. Moreover, $\mathcal{U}$, $\mathcal{Q}_\mu$ and $\mathcal{P}_{\mu\nu}$ represent respectively the nonlocal energy density, the nonlocal energy flux and the nonlocal anisotropic stress.

For a static spherically symmetric space-time, the nonlocal energy flux and nonlocal anisotropic stress become:
\begin{eqnarray}
&&\mathcal{Q}_\mu=0,\\
&&\mathcal{P}_{\mu\nu}=\mathcal{P}\left(r_\mu r_\nu-\frac{1}{3}h_{\mu\nu}\right),\label{bw6}
\end{eqnarray}
with $r^{\mu}$ being a unit radial vector and $h_{\mu\nu}=g_{\mu\nu}+u_\mu u_\nu$, such that $g_{\mu\nu}$ is the metric and $u_\mu$ is the four-velocity.

\subsection{Static structure equations}

With the aim of describing the properties of a spherically symmetric static fluid distribution, it is considered the line element in Schwarzschild coordinates:
\begin{equation}\label{bwg1}
ds^{2}=-A^{2}(r)dt^{2}+B^{2}(r)dr^{2}+r^{2}(d\theta^{2}+\sin^{2}\theta d\phi^{2}),
\end{equation}
with $A(r)$ and $B(r)$ being metric potentials.

The nonzero components of the Einstein's field equations on the brane for the metric above are:
\begin{eqnarray}
&&\frac{1}{r^2}-\frac{1}{r^2B^2}\left[1-2r\frac{B'}{B}\right]=k^{2}\rho^{\rm eff},\label{bwg2}\\
&&\frac{1}{r^2}-\frac{1}{r^2B^2}\left[1+2r\frac{A'}{A}\right]=-k^{2}p^{\rm eff}_r,\label{bwg3}\\
&&\frac{1}{B^{2}}\left[\frac{A''}{A}+\frac{A'}{rA}-\frac{A'B'}{AB}-\frac{B'}{rB}\right]=k^2\,p^{\rm eff}_t.
\end{eqnarray}
with primes denoting radial derivatives. The functions $\rho^{\rm eff}$, $p^{\rm eff}_r$ and $p^{\rm eff}_t$ are given by the equalities:
\begin{eqnarray}
&&\rho^{\rm eff}=\rho\left(1+\frac{\rho}{2\lambda}\right)+\frac{6\,\mathcal{U}}{k^4\,\lambda},\\
&&p^{\rm eff}_r=p+\frac{\rho}{2\lambda}\left(\rho+2p\right)+\frac{2\,\mathcal{U}}{k^4\,\lambda}+\frac{4\,\mathcal{P}}{k^4\,\lambda},\\
&&p^{\rm eff}_t=p+\frac{\rho}{2\lambda}\left(\rho+2p\right)+\frac{2\,\mathcal{U}}{k^4\,\lambda}-\frac{2\,\mathcal{P}}{k^4\,\lambda}.
\end{eqnarray}
From $p^{\rm eff}_r\neq p^{\rm eff}_t$, it can be understood that the effects of the extra dimensions produce anisotropy in the fluid contained in the star.

For our purposes, it will also be of great importance to know the covariant derivative of both the energy-momentum tensor and effective energy-momentum tensor on the brane. The conservative equations $\nabla^{\nu}T_{\mu\nu}=0$ and $\nabla^{\nu}T_{\mu\nu}^{\rm eff}=0$ for the metric (\ref{bwg1}) read, respectively
\begin{eqnarray}
&&\hspace{-0.5cm}p'+\frac{A'}{A}(\rho+p)=0,\label{bwg4} \\ 
&&\hspace{-0.5cm}\mathcal{U}'+\frac{2\,A'}{A}(2\mathcal{U}+\mathcal{P})+2\mathcal{P}'+\frac{6\mathcal{P}}{r}=-\frac{k^{4}}{2}\rho'(\rho+p). \label{bwg5}
\end{eqnarray}
We remark here, for the sake of completeness, that $\rho(r)$, $p(r)$, $\mathcal{P}(r)$ and $\mathcal{U}(r)$, as well as $A(r)$ and $B(r)$,  depend on the radial coordinate only. 

\section{Gravastar in the Braneworld}\label{sec:bwg}

 \subsection{General remarks}

The static structure of the gravastar under study is envisaged in the following form: the interior of the object is surrounded by a thin shell of ultra-relativistic fluid, while the outer space-time is described by a vacuum exterior solution. The three regions aforementioned are structured considering the following equations of state:
\begin{itemize}
\item Interior: \hspace{0.3cm}$0\leq r<r_1$; \hspace{0.3cm}$p=-\rho$,
\item Shell:    \hspace{0.55cm}$r_1<r<r_2$;  \hspace{0.3cm}$p=\rho$,
\item Exterior: \hspace{0.0cm}$r_2<r$;       \hspace{1.1cm}$p=\rho$=0,
\end{itemize}
with $r_1$ and $r_2$ being the interior and exterior radii of the gravastar, respectively.

In addition, with the aim of comparing our results to those obtained by Mazur and Mottola \cite{mazur/2004}, we consider $\mathcal{U}=0$ in both the interior and shell of the gravastar.  Moreover, we regard that the outer space-time is described by a Schwarzschild vacuum solution. 

\subsection{Interior of the gravastar}

First of all, it is important to say that in the interior region of the gravastar, from Eq.~(\ref{bwg4}), $p=-\rho=-\rho_0={\rm cte.}$ .

Now, following \cite{mazur/2004}, we define the potential metric $B^2$ in such a form that:
\begin{equation}\label{def_b}
B^{-2}=\left(1-I^{2}_{0}r^2\right),
\end{equation}
with $I_0$ being a constant. Considering Eq.~(\ref{def_b}) in  Eq.~(\ref{bwg2}), one obtains that $I_0$ and $\rho_0$ are connected through
\begin{equation}\label{eq_ho}
I_0=\sqrt{\frac{k^2}{3}\left(\rho_0+\frac{\rho_0^2}{2\lambda}\right)}.
\end{equation}

On the other hand, by replacing the nonlocal pressure

\begin{equation}\label{P_1}
\mathcal{P}=-\left(\frac{k^2\lambda}{2r}\right)\frac{k_1}{1-k_1}\frac{B'}{B^3},
\end{equation}
with constant $k_1$, in the sum of the Eqs.~(\ref{bwg2}) and (\ref{bwg3}), we can obtain an equation that relates the two potential metrics $A$ and $B$ as
\begin{equation}\label{eq_a_b_k}
A=k_2\,B^{-\frac{1}{1-k_1}},
\end{equation}
where $k_2$ represents an integration constant. From Eq.~(\ref{eq_a_b_k}), for $k_1=0$ we determine that the interior region is described by the de Sitter metric.

We note that another form for the function $\mathcal{P}$ can be found by integrating Eq.~(\ref{bwg5}), resulting in
\begin{equation}\label{eq_p_k1}
\mathcal{P}=\frac{k^{3}_{3}}{A\,r^3},
\end{equation}
with constant $k_3$.

Evidently, the functions presented in (\ref{P_1}) and (\ref{eq_p_k1}) must be equal, thus, we find that the constants $k_1$ and $k_3$ are related through:
\begin{equation}\label{k_1_and_k_3}
k_1=\frac{k_{3}^{3}}{k_{3}^{3}-k^2\lambda\,r^3I_0^2\left(1-r^2I_0^2\right)^{\frac{1}{2(1-k_1)}}}.
\end{equation}
In order to obtain regular solutions we need to have:
\begin{equation}
k^2\lambda\,r^3I_0^2\left(1-r^2I_0^2\right)^{\frac{1}{2(1-k_1)}}\neq k_{3}^{3}.
\end{equation}
From Eq.~(\ref{k_1_and_k_3}), note that if $k_3=0$ then $k_1=0$ and if $\lambda\to\infty$ then $k_1=0$ ($k_3=0$). In both cases we derive
$\mathcal{P}=0$, indicating that the Mazur-Mottola case \cite{mazur/2004} is obtained.

\subsection{Shell}

Such as aforementioned, we consider that the pressure and energy density of the fluid contained in the shell are related through  $p=\rho$.

In order to determine $\mathcal{P}$, we replace (\ref{bwg4}) in (\ref{bwg5}) yielding:
\begin{equation}\label{p_rho2}
\mathcal{P}=-\frac{k^4}{3}\rho^2.
\end{equation}

Now, as well as it is considered by Mazur and Mottola \cite{mazur/2004}, let us introduce a dimensionless variable $\xi$ as
\begin{equation}\label{anzat}
\xi=k^2r^2\rho.
\end{equation}

Replacing Eq.~(\ref{anzat}) together with Eq.~(\ref{p_rho2}) and Eq.~(\ref{bwg4}) into Eqs.~(\ref{bwg2}) and (\ref{bwg3}), respectively, these can be rewriten as:
\begin{eqnarray}
&&\frac{dr}{r}=\frac{d\left(B^{-2}\right)}{1-\frac{1}{B^2}-\xi-\frac{\xi^2}{2\lambda\,k^2r^2}},\label{eq_B_r}\\
&&\frac{d\left(B^{-2}\right)}{B^{-2}}=-\left[\frac{1-\frac{1}{B^2}-\xi-\frac{\xi^2}{2\lambda\,k^2r^2}}{1-\frac{3}{B^2}+\xi+\frac{\xi^2}{6\lambda\,k^2r^2}}\right]\frac{d\xi}{\xi}.\label{eq_B_2}
\end{eqnarray}

We can note that it is difficult to obtain analytical solutions from these field equations. Nevertheless, this can be achieved by taking into account that in the thin shell limit, $0<B^{-2}<<1$. Under this limit, the integration of Eq.~(\ref{eq_B_2}) yields
\begin{equation}\label{eq_B_2_aprox}
B^{-2}\simeq\epsilon\frac{\left[\frac{\xi^2}{6\lambda\,k^2r^2}+\xi+1\right]^2}{\xi},
\end{equation}
where $\epsilon$ is an integration constant. Moreover, since $B^{-2}<<1$ we need that $\epsilon<<1$ also.

Finally, by making use of Eqs.~(\ref{eq_B_r}) and (\ref{eq_B_2_aprox}), we obtain:
\begin{equation}\label{dr_dw_w2}
dr\simeq-\epsilon\,r\left[\frac{\xi^2}{6\lambda\,k^2r^2}+\xi+1\right]\frac{d\xi}{\xi^2}.
\end{equation}

\subsection{Exterior of the gravastar}

For this region, we consider that $\mathcal{U}=\mathcal{P}=0$. This ensures that the exterior space-time is described by the Schwarzschild vacuum metric. Thus, the gravastar outer space-time is depicted by the line element:
\begin{equation}
ds^2=-Fdt^2+F^{-1}dr^2+r^2\left(d\theta^2+\sin^2\theta d\phi^2\right),
\end{equation}
with
\begin{equation}\label{Eq_F}
F=1-\frac{2MG}{r},
\end{equation}
where $M$ represents the total mass of the gravastar.

\section{Junction conditions}

As previously shown, the study of gravastars involves an inner region and an outer region separated by a shell of matter which we shall denominate $\Sigma$. In order to realign the physical and geometric quantities of the inner and outer regions with the magnitudes of surface, we will use the conditions of continuity of Israel-Darmois \cite{Israel/1966,Darmois/1927}. They say that the metric coefficients are continuous in $\Sigma$ ($r=R$), but their derivatives are not continuous at this point. 

It is possible to determine the surface energy-momentum tensor with the help of the Lanczos equation \cite{Lanczos/1924}:
\begin{equation}\label{surface_TEM}
{\cal S}^{i}_{j}=\frac{1}{8\pi}\left(\kappa^{i}_{j}-\delta^{i}_{j}\kappa^{k}_{k}\right),
\end{equation}
with the Latin indexes running as $i, j=t, \theta, \phi$. The factor $\kappa_{ij}$ depicts the discontinuity in the extrinsic curvature $K_{ij}$, with $\kappa_{ij}=K^{+}_{ij}-K^{-}_{ij}$, where the signs $-$ and $+$ correspond respectively to the interior and exterior regions. The extrinsic curvature is defined by:
\begin{equation}
K_{ij}^{\pm}=-n_{\beta}^{\pm}\left(\partial_{j}e_{i}^{\beta}+\Gamma_{\mu\nu}^{\beta}e_{i}^{\mu}e_{j}^{\nu}\right),
\end{equation}
with $e_{i}^{\mu}=\frac{\partial x^{\mu}}{\partial\zeta^{i}}$, where $\zeta^{i}$ represents the coordinate on the shell, $n_{\beta}^{\pm}$ depicts the normal vector to the surface and $\Gamma^{\beta}_{\mu\nu}$ refers to the Christoffel symbols.

Once considered ${\cal S}^{i}_{j}={\rm diag}(\sigma,-v,-v)$, with $\sigma$ and $v$ being respectively the surface energy density  and the surface pressure, the Lanczos equation can be placed on the form:
\begin{eqnarray}
&&\sigma=-\frac{1}{4\pi}\kappa^{\theta}_{\theta},\label{sigma}\\
&&v=\frac{1}{8\pi}\left(\kappa^{t}_{t}+\kappa^{\theta}_{\theta}\right).
\end{eqnarray}

Using Eqs.~(\ref{def_b}), (\ref{Eq_F}) and (\ref{sigma}), the thin shell mass can be found using the equality:
\begin{equation}\label{def_ms}
m_s=4\pi\,R^2\sigma=-R\sqrt{1-\frac{2M}{R}}+R\sqrt{1-I^{2}_{0}R^2}.
\end{equation}

Considering (\ref{eq_ho}), from Eq.~(\ref{def_ms}) we obtain that the total mass of the gravastar is given by:
\begin{equation}\label{tota_mass}
M=\frac{R}{2}-\frac{R}{2}\left[\sqrt{1-\frac{k^2R^2}{3}\left(\rho_0+\frac{\rho_0^2}{2\lambda}\right)}-\frac{m_s}{R}\right]^2.
\end{equation}
It is important to mention that if we consider $\lambda\to\infty$, Eq.~(\ref{tota_mass}) is reduced to the same equation than those found in  \cite{das/2017} and \cite{Banerjee/2016}, in their particular cases.

\section{Some physical features of the model}

It is important to highlight that the shell of the gravastar is limited by the interfaces $R_1=R$ e $R_2=R+\epsilon$, thus connecting the inner space-time with the outer space-time. In order to analyze the principal physical characteristics of the matter in the shell, some definitions used by Mazur and Mottola \cite{mazur/2004} are considered throughout this section.

\subsection{The proper thickness of the shell}

The proper thickness of the shell is determined from
\begin{equation}
\ell=\int_{R_1}^{R_2} Bdr.
\end{equation}
Using (\ref{eq_B_2_aprox}) and (\ref{dr_dw_w2}) in the equation above, it becomes
\begin{equation}\label{shell}
\ell\simeq\epsilon^{1/2}r\int_{\xi_1}^{\xi_2}\xi^{-3/2}d\xi\simeq2\epsilon^{3/2}R,
\end{equation}
where we notice that $\ell$ is very small in relation to $R$.  This last result is the same found by Mazur and Mottola \cite{mazur/2004}. In this way we understand that the 5D bulk has no effects on the proper thickness of the shell.

\subsection{Energy within the shell}

The energy inside the thin shell is determined as
\begin{equation}\label{energy_shell}
{\cal E}=4\pi\int_{R_1}^{R_2}\rho^{\rm eff}r^2dr.
\end{equation}
Considering Eqs.~(\ref{anzat}) and (\ref{dr_dw_w2}) in Eq.~(\ref{energy_shell}), we obtain
\begin{equation}
{\cal E}=\frac{4\pi\epsilon R}{k^2}\int_{\xi_2}^{\xi_1}\left[1+\frac{\xi}{2\lambda k^2r^2}\right]\left[1+\frac{1}{\xi}+\frac{\xi}{6\lambda k^2r^2}\right]d\xi,
\end{equation}
which integrated yields
\begin{eqnarray}
\fl&&{\cal E}=\frac{4\pi\epsilon R}{k^2}\left[\ln(\xi)+\frac{1}{9}\left[9\left(\frac{1}{2\lambda R^2k^2}+1\right)\xi+6\left(\frac{1}{2\lambda R^2k^2}\right)\xi^2+\left(\frac{1}{2\lambda R^2k^2}\right)^2\xi^3\right]\right]_{\xi_2}^{\xi_1}.
\end{eqnarray}
Once $\epsilon<<1$, the energy ${\cal E}$ until  first order of $\epsilon$ is given by
\begin{eqnarray}\label{resulting_eq}
{\cal E}\simeq\frac{8\pi\epsilon^2 R}{k^2}\left[1+\left(\frac{1}{2}+\frac{2\xi_2}{3}\right)\left(\frac{1}{2\lambda R^2k^2}\right)+\frac{\xi_2^2}{6}\left(\frac{1}{2\lambda R^2k^2}\right)^2\right].
\end{eqnarray}

The resulting nonlinear equation provides information about the energy within the shell. It indicates that the energy is directly proportional to $\epsilon^2$. Note that the energy in the shell, Eq.~(\ref{resulting_eq}), is larger than the one derived by Mazur and Mottola \cite{mazur/2004}. This is due to the effects of the 5D bulk on the brane which helps to increase the energy within the shell. In comparison with an alternative gravity gravastar, such as the one found in $f(R,T)$ gravity for instance, it can be noted a stronger dependence of the energy with $\epsilon$ in the braneworld model with the obtained by $f(R,T)$ gravity \cite{das/2017}.

\subsection{Entropy of the shell}

We calculate the entropy in the shell as
\begin{equation}\label{entropy}
S=4\pi\int_{R_1}^{R_2} s\,r^2Bdr,
\end{equation}
where $s$ represents the local specific entropy density, given by:
\begin{equation}\label{entropy2}
s=\alpha\frac{k_B}{\hbar}\sqrt{\frac{p}{2\pi G}},
\end{equation}
with $\alpha$ being a dimensionless constant, $k_B$ representing the Boltzmann constant and $\hbar=h/(2\pi)$ where $h$ is the Planck constant.

Considering Eqs.~(\ref{anzat}), (\ref{eq_B_2_aprox}) and (\ref{dr_dw_w2}) in (\ref{entropy}), (\ref{entropy}) becomes:
\begin{equation}\label{entropy3}
S\simeq\frac{\alpha k_B}{\hbar G}\epsilon^{1/2}R^2\ln\left(\frac{\xi_1}{\xi_2}\right).
\end{equation}

As in Ref.\cite{mazur/2004}, by taking into account that $\xi_1/\xi_2=1+{\cal O}(\epsilon)$ as well as  Eq.~(\ref{shell}), we have that Eq.~(\ref{entropy3}) yields:

\begin{equation}\label{entropy4}
S\simeq\frac{\alpha k_B R\ell}{2\hbar G}.
\end{equation}
From Eq.~(\ref{entropy4}), we can see that the entropy depends directly on the proper thickness of the shell. This result is equal to the one derived by Mazur and Mottola in \cite{mazur/2004}. This shows that the 5D bulk does not affect the entropy of the shell. Comparing with gravastars obtained in $f(R,T)$ gravity \cite{das/2017}, we note that the entropy has a substantial dependence on $\epsilon$ as $\epsilon^{3/2}$ while in $f(R,T)$ gravity, it goes approximately with $\epsilon^{1/2}$.

\section{Discussion}

With recent advances in gravitational wave observational astronomy, several studies of gravastars as possible sources of gravitational radiation have been made. Let us briefly review some of those important contributions.

In \cite{pani/2009}, it was deeply discussed how the presence or absence of an event horizon can produce qualitative differences in the gravitational waves emitted by ultracompact objects. In \cite{pani/2010}, it was shown that the gravitational sign emitted by a gravastar could provide a unique signature of the horizonless nature of such an object. In \cite{uchikata/2016} it was discussed how a measurement of the tidal deformability from the gravitational-wave detection of a compact-binary inspiral can be used to constrain gravastars. 

A further alternative to detect gravastars was proposed in \cite{kubo/2016}, through their gravitational lensing. Naturally, the recently reported results regarding the M87 BH shadow \cite{event_horizon_collaboration/2019} can also in near future help us to distinguish BHs and gravastars.

We have constructed in the present paper gravastar solutions in RSII. Besides \cite{das/2017,Banerjee/2016}, gravastars have also been constructed in alternative gravity theories in References \cite{bhar/2014,rahaman/2015}.

Alternative gravity theories have been mostly used to try to account for the cosmological dark sector of the universe. In fact, it is possible, through modified gravity, to describe the galactic and cosmological scales of the universe without dark matter and dark energy \cite{capozziello/2007,zlosnik/2007,nojiri/2006,joyce/2016,kase/2019,nojiri/2005,woodard/2007,cognola/2006,kase/2018,amendola/2007}. Although RSII has been proposed as an alternative to the hierarchy problem, it also works pretty well in cosmology. On this regard, besides \cite{ramirez/2004,holanda/2013,hebecker/2001,meehan/2014}, one can also check References \cite{nozari/2009,barros/2016}.

The investigation of gravastars in braneworld scenarios can give us some new insights on both the geometrical and physical features of these objects and the braneworld setup itself.

Particularly, here we have derived different gravastar physical parameters in the RSII, as Mazur and Mottola have done in standard gravity scenario. We noted that both the mass and shell energy are altered with respect to the Mazur-Mottola original results due to the presence of the brane tension. On the other hand, the proper thickness of the shell and the shell entropy are not altered due to RSII configuration.  The present model can be used to investigate the possible 5D effects in some physical phenomenal that arise in the study of gravastars. An extent of the present formalism can be used to analyze the ergoregion instability in rotating gravastars \cite{chirenti/2008,cardoso/2008}, since the projections on the brane could help the stability of gravastars against the ergoregion instability, such as they help the fluid pressure of compact stars to support more mass against the gravitational collapse \cite{la/2017,la/2015}.

Let us further analyse the physical features of our model. It should also be interesting to compare our results with the present literature in alternative gravity gravastars, such as the model presented in \cite{das/2017} within the $f(R,T)$ gravity \cite{harko/2011}, for which $f(R,T)$ is a general function of the Ricci scalar $R$ and trace of the energy-momentum tensor $T$.

The proper thickness $\ell$ of the shell found in our model is directly proportional to $R$, the radius of the inner shell. It is also proportional to the thickness of the shell as $\ell\sim\epsilon^{3/2}$. Since $\epsilon<<1$, the latter proportionality indicates that $l<<R$.

The entropy we have obtained is gradually increasing with respect to $\epsilon$, a result also obtained in $f(R,T)$ gravity \cite{das/2017}. On the other hand, the energy within the shell has a stronger dependence on $\epsilon$ as $\sim\epsilon^2$ when compared to the $f(R,T)$ gravity result, which reads  $\sim\epsilon$.  

To finalize we remark that our results retrieve the Mazur-Mottola results in the regime $\lambda^{-1}\rightarrow{\infty}$, as it is expected in braneworld features.

\section*{Acknowledgment}
PHRSM would like to thank Funda\c{c}\~ao de Amparo \`a Pesquisa do Estado de S\~ao Paulo (FAPESP), grant $2018/20689-7$. The authors thank FAPESP for financial support under the projects $2013/26258-4$.

\end{document}